\documentstyle [twocolumn,prc,aps,epsfig]{revtex}
\begin {document}
\twocolumn[\hsize\textwidth\columnwidth\hsize\csname @twocolumnfalse\endcsname
\title
{Enhancement of $\pi A \to \pi\pi A$ Threshold Cross Sections
by In-Medium $\pi\pi$ Final State Interactions}
\author
{R. Rapp$^{1}$, J.W. Durso$^{2,3}$, Z. Aouissat$^4$, G. Chanfray$^5$, 
 O. Krehl$^2$, P. Schuck$^6$, J. Speth$^2$ and J. Wambach$^{4}$}
 
\address
{1) Department of Physics and Astronomy, SUNY at Stony Brook, 
    Stony Brook NY 11794-3800, U.S.A. \\
 2) Physics Department, Mount Holyoke College, South Hadley,
    MA 01075, U.S.A.\\
 3) Institut f\"{u}r Kernphysik, Forschungszentrum
     J\"{u}lich GmbH, D-52425 J\"{u}lich, Germany \\
 4) Institut f\"ur Kernphysik, Schlo{\ss}gartenstr.~9, TU-Darmstadt, 
    D-64289 Darmstadt, Germany \\
 5) IPN, IN2P3-CNRS/Universite Claude Bernard Lyon I, 
    43 Bd. du 11 Nov. 1918 F-69622 Villeurbanne Cedex, France \\
 6) ISN, IN2P3-CNRS/Universite Joseph Fourier, 53 Av. de Martyrs, 
    F-38026 Grenoble Cedex, France}
\date{\today} 
\maketitle
 
\begin{abstract} 

We address the problem of pion production in low energy $\pi$-nucleus
collisions.  For the production mechanism we assume a simple model
consisting of a coherent sum of single pion exchange and the 
excitation---followed by the decay into two pions and a nucleon---of 
the $N^*(1440)$ resonance.  The production amplitude is modified by 
the final state interaction between the pions calculated using the 
chirally improved J\"ulich meson exchange model including 
the polarization of the nuclear medium by the pions.  The model
reproduces well the experimentally observed $\pi A \to \pi\pi A$ 
cross sections, especially the enhancement with increasing $A$ of the
$\pi^+\pi^-$ mass distribution in the threshold region.  

\end{abstract}
\vspace{0.6cm}
]


The past ten years have witnessed a considerable increase in our
understanding of the $\pi\pi$ interaction.  The success of chiral 
perturbation theory~\cite{GaLe,Don} in describing the near-threshold 
behavior of the $\pi\pi$ amplitude is a clear success for the
effective field theory approach to the problem.  Unfortunately,
at energies above a few hundred MeV, where effects of unitarity
become important, chiral perturbation theory becomes unwieldy and
other approaches must be used to obtain a good 
description of the free $\pi\pi$ scattering.  One such approach is
the meson exchange model developed by the 
J\"ulich group~\cite{Jpipi}, which
gives an excellent quantitative description of $\pi\pi$ and $\pi K$
scattering phases up to about 1.5 GeV total cm energy.  Another is
the inverse amplitude method---a variant of the $K$-matrix 
approach---of Oset et al.~\cite{OlOs}. Here we will consider
the application of the former to the problem of pion production by
pions on nuclei in order to investigate some of the predictions of
the model for the behavior of the $\pi\pi$ scattering amplitude in 
the presence of a nuclear medium. The latest version of this model, 
which we refer to as the chirally improved J\"ulich model~\cite{RDW96}, 
respects the constraints on the $S$-wave scattering length imposed by
chiral symmetry while maintaining the quality of fit to the free
$\pi\pi$ scattering data.  We briefly summarize here the main 
features of the model and present some of the results of the model
for $\pi\pi$  interactions in nuclear matter.  For details of the 
model we direct the reader to ref.~\cite{RDW96,RPhD}.


In the chirally improved J\"ulich model, the interaction is driven
by the exchange of $\rho$ mesons plus contact interactions, as in
the Weinberg lagrangian~\cite{We66}. This lagrangian is chirally 
symmetric in the massless pion limit.  The Born approximation for 
this interaction is used as the potential in a three-dimensional 
scattering equation of the Blankenbecler-Sugar form~\cite{BbS}.  
The solution
of the scattering equation destroys chiral symmetry through both the
partial summation of diagrams and the use of form factors, which are
needed to ensure convergence of the integral equation.  However, the
off-shell behavior of the potential is prescribed in such a way as
to preserve the scattering length constraint required by chiral
symmetry.

\begin{figure}
\vspace{-0.6cm}
\centerline{\epsfig{file= 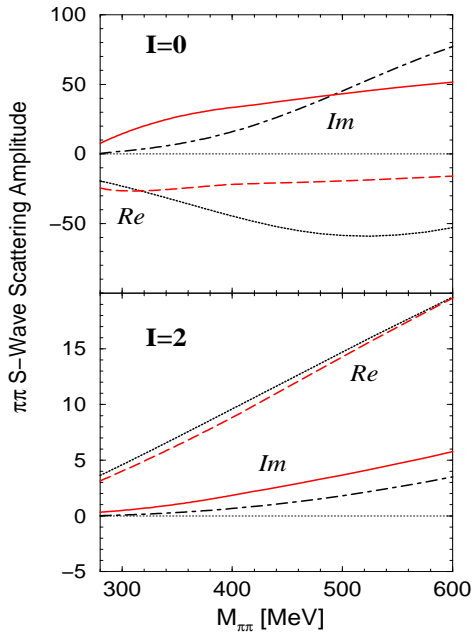,width=8cm,height=10cm}}
\vspace{-0.3cm}
\caption{Dimensionless S-Wave $\pi\pi$ scattering amplitudes
for total isospin 0 (upper panel) and 2 (lower panel); the dotted 
and dashed-dotted lines correspond to the real and imaginary parts, 
respectively, in free space, and the long-dashed and full lines 
correspond to the real and imaginary parts, respectively, at a nuclear
density of $\rho=0.5\rho_0$.}
\label{Mpipi}
\end{figure}

The motivation to ``chirally improve" the original J\"ulich model
resulted from studies of the behavior of the $\pi\pi$ interaction in 
nuclear matter as a function of density~\cite{pipidens}. In these 
studies the polarization of the medium by the pion, principally through
the production of nucleon-hole and $\Delta$-hole configurations,
significantly reshaped the $\pi\pi$ scattering amplitude.  It was 
observed that models in which the potential term is not 
chirally symmetric in the limit of zero pion mass led to a rapid
increase in attraction between the pions as a function of density,
resulting ultimately in spontaneous $S$-wave pion pair condensation at 
densities about that of normal nuclear matter.  Further 
studies~\cite{ARCSW,RDW96,RPhD}
showed that the onset of this instability could be shifted to 
higher densities if the $\pi\pi$ potential term were required to
satisfy the low-energy chiral constraints, or if the amplitude
calculated from the scattering equation could be forced to obey
the same constraints.  However, the build up of attractive strength
in the threshold and sub-threshold region persisted. 
In fig.~\ref{Mpipi} we show the $S$-wave 
$\pi\pi$ scattering amplitudes at low energies 
in free space and at half nuclear matter density using the chirally 
improved J\"ulich model from ref.~\cite{RDW96}, which will be employed 
in the following. Whereas the isospin $I=0$ amplitudes (upper panel) 
undergo substantial reshaping in the medium, the $I=2$ channel (lower 
panel) is hardly affected. In the $I=0$ channel, the combination of 
the attractive $\pi\pi$ interaction and the softening of the in-medium 
single-pion dispersion relation leads to a substantial accumulation 
of strength in the imaginary part of the in-medium amplitude 
(notice, however, that the real part is actually only enhanced for 
$M_{\pi\pi}\le 300$~MeV). In contrast, the purely repulsive 
interaction in the $I=2$ channel prevents essential modifications.  
Since the in-medium scalar-isoscalar $\pi\pi$ amplitude is the 
main mediator of the intermediate range attraction in the 
nucleon-nucleon interaction, it is of prime importance for e.g. 
nuclear saturation. In fact, in refs.~\cite{RDW97,RMDB} the use of the 
chirally improved J\"ulich model in the Bonn Potential for the $NN$ 
interaction~\cite{MHE} has been shown to be compatible with the empirical 
nuclear saturation point  once medium modifications
in the vector mesons exchanges (providing short-range repulsion) are
also included. Here we are interested in a more direct assessment 
of the in-medium $\pi\pi$ $S$-wave correlations.

A challenging testbed for the in-medium modifications of the 
two-pion interaction discussed above has been set by recent 
experimental data taken by the CHAOS 
collaboration~\cite{chaos96,chaos98} at TRIUMF. An incoming pion 
beam of nominal kinetic energy $T_\pi$=282.7~MeV was directed at 
various nuclear targets, and two outgoing pions were detected in 
coincidence. With increasing nuclear mass number, the corresponding 
invariant mass spectra show a strong enhancement of $\pi^+\pi^-$ 
pairs just above the two-pion threshold of $M_{\pi\pi}=2m_\pi$, 
whereas only very minor variations are observed for 
$\pi^+\pi^+$ pairs.

That the $\pi^+\pi^-$ strength in the scalar-isoscalar channel 
in a nuclear medium can be strongly enhanced at threshold was 
predicted in ref.~\cite{SNC}, based on inclusive data from the CHAOS
collaboration~\cite{chaos87}. A first indication that this threshold
enhancement can account for the measured $(\pi^+,\pi^+\pi^-)$ cross
section at low invariant mass was obtained from a schematic calculation
in ref.~\cite{Sch98}. 
A clue to the possible underlying mechanism for this 
dramatic effect is provided by the isospin decomposition of 
the charged states. For $S$-waves one has   
\begin{eqnarray} 
{\cal M}_{\pi^+\pi^-} & = & \frac{2}{3} {\cal M}^{I=0} +
\frac{1}{3} {\cal M}^{I=2} 
\nonumber\\
{\cal M}_{\pi^+\pi^+} & = & {\cal M}^{I=2} \ , 
\end{eqnarray}
showing that while the $\pi^+\pi^+$ channel is pure isotensor, 
the $\pi^+\pi^-$ channel is predominantly isoscalar. 
Thus, the in-medium $\pi\pi$ amplitudes~\cite{RDW96} shown in 
fig.~\ref{Mpipi} clearly exhibit desirable
features: little modification of the scalar-isotensor amplitude, and a 
significant low energy enhancement of the scalar-isoscalar amplitude.  
To quantify the comparison, the actual cross sections for the 
$\pi A \to \pi\pi A$ production process must be calculated, 
including the experimental acceptance of the CHAOS spectrometer, 
which is crucial to reproducing the precise shapes of the observed 
spectra.  
        
Since our emphasis here is on possible medium effects in the 
$\pi\pi$ interaction, we will treat the pion production process
in a simplified manner. First, we assume that it always proceeds as
an elementary process on a single nucleon. Second, we account only  
for the two most important contributions which, according to 
refs.~\cite{Os85,chaos98},  are the one-pion exchange (OPE) reaction 
(contributing to both isoscalar and isotensor channel) and the 
$N^{*}(1440)$ resonance formation (leading to isoscalar $\pi\pi$ states
only). 

Let us first discuss the OPE contribution.  The corresponding 
4-differential cross section is given by  
\begin{eqnarray}
\frac{d^4\sigma_{\pi N\to \pi\pi N}^{OPE}}
{dM_{\pi\pi} dt d\cos\theta_{cm} d\phi_{cm}}= 
\frac{M_{\pi\pi}^2}{16\pi^2 p_{lab}^2} \ 
\frac{IF f_{\pi NN}^2}{4\pi m_\pi^2} 
\nonumber\\
 \ \times |t|  
\ \frac{|{\cal M}_{\pi\pi}(M_{\pi\pi},q,q',\cos\theta_{cm})|^2}
{32\pi M_{\pi\pi}^2} \ q' \ D_\pi(t)^2,  
\label{dsigope}
\end{eqnarray} 
where $M_{\pi\pi}$ is the two-pion invariant mass, 
$D_\pi(t)=[t-m_\pi^2]^{-1}$ the 
propagator of the exchanged pion, $p_{lab}$ the incoming 
pion laboratory momentum, and
$q,q',\phi_{cm},\theta_{cm}$ the in/outgoing pion momentum and 
their relative angles in the two-pion cms ($f_{\pi NN}$=1.01 and $IF$=2
for charged pion exchange). 
  Azimuthal symmetry in the $\pi\pi$ interaction 
makes the $\phi_{cm}$ integration trivial. Furthermore, since our focus 
will be on the near threshold region, the $S$-Wave contribution is 
expected to be dominant. This is supported  by  experimental
angular distributions in $\cos \theta_{cm}$ which do not show any trace 
of other than (isotropic) $S$-wave components~\cite{chaos96}. 
We will thus neglect $P$- 
and higher partial waves. The invariant mass spectra are then 
given by 
\begin{equation}
\frac{d\sigma^{OPE}} {dM_{\pi\pi}}  = 
4 \pi \int\limits_{t_{min}}^{t_{max}} dt  \
 \frac{d^2\sigma_{\pi N\to \pi\pi N}^{OPE}}
{dM_{\pi\pi} dt}  \ Acc(t,M_{\pi\pi},s_{tot}) \ .  
\label{ope}
\end{equation} 
For a meaningful comparison with the experimental spectra it is essential
to include the acceptance of the CHAOS spectrometer,  represented by an
'acceptance factor' $Acc(t,M_{\pi\pi},s_{tot})$ in eq.~(\ref{ope}). 
We evaluate it by Monte Carlo techniques: in the two-pion CMS we randomly  
generate pairs of angles $(\phi_{cm}, \cos\theta_{cm})$, uniformly
distributed over the sphere due to the $S$-wave nature of the 
$\pi\pi$-amplitude. (To include higher partial waves one would need to 
weight the polar angles with the corresponding angular distribution
obtained from the total amplitude for the particular $M_{\pi\pi},q,q'$.)
For given $t,M_{\pi\pi},s_{tot}$ one can 
then determine the total momentum $\vec P$ of the pion pair in the 
laboratory frame and apply the corresponding Lorentz boost to transform
the two generated tracks into the lab system. In the latter, 
the experimental acceptance cuts can then be readily applied,  
{\it i.e.} $\Phi_{\pi\pi}=0^o\pm 7$ or $\Phi_{\pi\pi}=180^o\pm 7$ for 
the out-of-reaction plane opening angle of the pion pair and 
$\Theta_{\pi^\pm}=10^o-170^o$ for the in-plane angle of both 
single pion tracks accounting for the dead regions around 
the beam-direction. At fixed $t,M_{\pi\pi}$ and $s_{tot}$ the 
acceptance probability is then determined as $Acc=N_{acc}/N_{tot}$, 
with $N_{tot}$ the number of trials and $N_{acc}$ the number of 
events that fall into the experimental acceptance.

In order to assess the $N^*(1440)$ 
contribution we model the two-pion production 
amplitude as proceeding via an intermediate scalar-isoscalar resonance
$\tilde\sigma$ of invariant mass $M_{\pi\pi}$, which subsequently 
decays into two pions,  
i.e $\pi N \to N^*(1440) \to \tilde\sigma N \to \pi\pi N$.   
Since in the J\"ulich model there is no low-lying genuine $\sigma$ 
resonance (the `$\sigma(550)$' being chiefly generated through attractive
$t$-channel $\rho$ exchange, and the $f_0(980)$ being realized as a 
$K\bar K$ bound state), we identify the $\tilde\sigma$ with the 
$\epsilon$ resonance located at around $M_{\pi\pi}\simeq$1400~MeV.
Its coupling to $\pi\pi$ states is well determined by a satisfactory fit 
to the $\delta^{00}_{\pi\pi}$ phase shifts above 1~GeV~\cite{RDW96}. 
For the $N^{*}(1440)N\epsilon$ vertex we assume scalar coupling, 
\begin{equation} 
{\cal L}_{N^*N\epsilon}= g_{N^*N\epsilon} \ \Psi_{N^*}^\dagger \ 
\epsilon \ \Psi_N + \ h.c.
\end{equation}
and estimate the corresponding coupling constant $g_{N^*N\epsilon}$ 
from the experimentally measured branching ratio of 5-10\% for 
$N^*(1440)\to N(\pi\pi)_{S-wave}^{I=0}$~\cite{PDG96}. Similarly, for 
the entrance channel, we employ the usual interaction vertex
\begin{equation}
{\cal L}_{N^*N\pi}= \frac{f_{N^*N\pi}}{m_\pi} \ \Psi_{N^*}^\dagger \
 \ \vec\sigma \cdot \vec q \ \tau_a \pi_a  \ \Psi_N + \ h.c. \ ,
\end{equation} 
and adjust the $N^*N\pi$ coupling constant to the branching ratio  
for $N^*(1440)\to N\pi$ of 60-70\%~\cite{PDG96}. In Born 
approximation, the cross section for 
$\pi N\to N^*(1440)\to \epsilon N$ with an $\epsilon$ of invariant 
mass $M_{\pi\pi}$ and 3-momentum $P$ is then obtained as
\begin{equation} 
\sigma_{\pi N\to N^*\to \epsilon N}(s_{tot},M_{\pi\pi})=
\frac{\overline{|{\cal M}_{N^*\epsilon}|^2} M_N}{8\pi p_{lab}^2} 
\int\limits_{P_{min}}^{P_{max}} dP \frac{P}{E_\epsilon(P)}
\label{sigeps}
\end{equation} 
with the spin averaged invariant matrix element in the $\pi N$ cms 
\begin{equation}
\overline{|{\cal M}_{N^*\epsilon}|^2}=\frac{IF}{2}  
\frac{f_{N^*N\pi}^2}{m_\pi^2} 
g_{N^*N\epsilon}^2 \frac{q_{in}^2}{|\sqrt{s_{tot}}-M_{N^*}-
{\rm i} \Gamma_{N^*}/2|^2} \ .  
\end{equation} 
The decay of the $\epsilon$ is now included by folding the cross
section eq.~(\ref{sigeps}) over the fully dressed $\epsilon$
spectral function which automatically accounts for the final
state interaction of the pion pair~\cite{ARCSW}:
\begin{eqnarray}
\sigma_{\pi N\to N^*\to \pi\pi N}(s_{tot})= \hspace{4cm}  
\nonumber\\
\int 
\frac{dM_{\pi\pi}  M_{\pi\pi}}{\pi} \ A_\epsilon(M_{\pi\pi}) \ 
 \sigma_{\pi N\to N^*\to \epsilon N}(s_{tot},M_{\pi\pi}) \ . 
\end{eqnarray} 
with 
\begin{eqnarray}
A_\epsilon(M_{\pi\pi}) & = & (D_\epsilon^0(M_{\pi\pi}))^2 \  
\pi \ v_{\epsilon\pi\pi}^2 \ q' \ M_{\pi\pi} \ 
\nonumber\\ 
 & & \quad  \times \left[ 1-\pi\ q' \ M_{\pi\pi} \ 
{\rm Im} {\cal M}_{\pi\pi}^{00}(M_{\pi\pi}) \right] \ ,
\end{eqnarray}
where $v_{\epsilon\pi\pi}$ denotes the ${\epsilon\pi\pi}$ vertex
function and ${\cal M}_{\pi\pi}^{00}$ the full $\pi\pi$ scattering
amplitude in the $JI=00$ channel.   
After a variable transformation, $dP=E_\epsilon(P)/(2M_NP) dt$, and
inclusion of the experimental acceptance, we finally obtain the differential
mass spectrum  
\begin{equation}
\frac{d\sigma^{N^*}}{dM_{\pi\pi}}=
\frac{\overline{|{\cal M}_{N^*\epsilon}|^2} M_{\pi\pi} 
A_\epsilon(M_{\pi\pi})} {16 \pi^2 p_{lab}^2} 
\int\limits_{t_{min}}^{t_{max}} dt \  Acc(t,M_{\pi\pi},s_{tot})  . 
\label{n*}
\end{equation}

The cross-section we wish to calculate requires  
a coherent sum of the $OPE$ and $N^*(1440)$  {\em amplitudes}. The
result of this yields the $M_{\pi\pi}$ differential cross section
\begin{eqnarray} 
\frac{d\sigma^{OPE+N^*}}{dM_{\pi\pi}}  =  \frac{q'}{16\pi^3 p_{lab}^2} 
\int\limits_{t_{min}}^{t_{max}} dt \ Acc(t,M_{\pi\pi},s_{tot})  
\nonumber \\
\times  |{\cal M}^{OPE} + {\cal M}^{N^*}|^2,
\end{eqnarray} 
where the amplitudes ${\cal M}^{OPE}$ and ${\cal M}^{N^*}$ can  be read 
off by comparison with  eqs.~(\ref{ope}) and (\ref{n*}), respectively.     

\begin{figure}
\centerline{\epsfig{file= 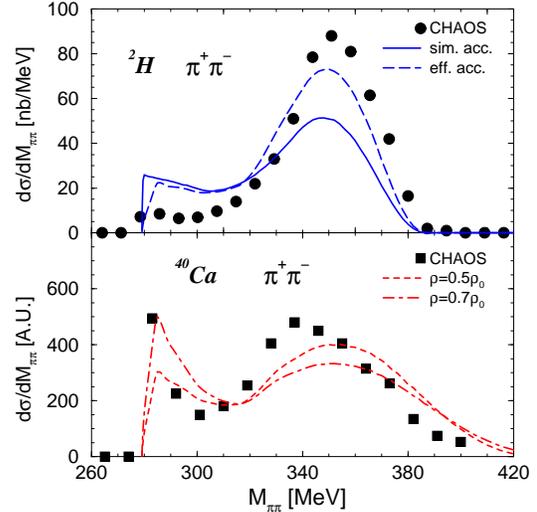, width=8cm, angle=-90}}
\vspace{0.3cm}
\caption{Results for pion production cross sections in the $\pi^+\pi^-$
channel using the chirally improved J\"ulich $\pi\pi$ interaction from
ref.~\protect\cite{RDW96};
upper panel: on the deuteron target using the Monte Carlo-simulated
acceptance (full lines) and the parameterized, `effective' acceptance;
lower panel: on Calcium, using the effective acceptance, for two different
densities. The $^2H$ and $^{40}Ca$ data are from ref.~\protect\cite{chaos98}
and \protect\cite{chaos96}, respectively.}
\label{+-}
\end{figure}

\begin{figure}
\centerline{\epsfig{file= 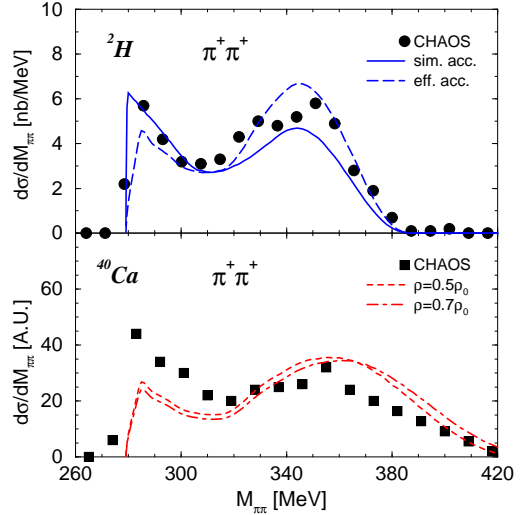, width=8cm, angle=-90}}
\vspace{0.3cm}
\caption{Results for pion production cross sections in the $\pi^+\pi^+$
channel using the chirally improved J\"ulich $\pi\pi$ interaction;
identical line identification as in fig.~\ref{+-}.}
\label{++}
\end{figure}

In the upper panels of figs.~\ref{+-} and \ref{++} our results for
deuterium target are displayed. The cross sections are calculated 
in absolute units using the free $\pi\pi$ interaction.  
The full lines correspond to a calculation with the Monte 
Carlo-simulated acceptance as described above.  
In the $\pi^+\pi^-$ channel the theoretical curves somewhat 
overestimate the data 
for $M_{\pi\pi}\le 320$~MeV, and underestimate them for 
$M_{\pi\pi}\ge 350$~MeV, a feature that is also shared by the model
calculations shown in ref.~\cite{chaos98}; in the $\pi^+\pi^+$ channel
the agreement is quite satisfactory. To check our determination of the 
experimental acceptance we also performed calculations with an 
`effective acceptance' (dashed lines in the upper panels); this 
was extracted from a fit to a phase space simulation given in 
ref.~\cite{chaos98}, which, after dividing out the phase space, leads 
to a simple $M_{\pi\pi}$-dependent acceptance factor $Acc(M_{\pi\pi})$.  
Whereas the description of the $\pi^+\pi^-$ channel turns out to be 
somewhat improved, the opposite trend is found in the $\pi^+\pi^+$ 
channel. In general, both ways of implementing the experimental 
acceptance agree reasonably well---within 20 percent.  

When moving to finite nuclei, the effect of the nuclear Fermi motion
has to be included.  In principle this is very complicated, even
if everything about the location of the interaction within the nucleus
were known.  We adopt instead a simplified procedure that, as we will
see, can adequately account for the observed cross-sections.  First, we
assume that the production of the pion pairs occurs at a definite
density.  In principle one should then average over all momenta within
the Fermi sphere for that density, which would mean calculating the
cross-section for non-collinear collisions, and we should also account
for the effective mass of the nucleon as a function of momentum.
This is computationally tedious and, given the precision of the data,
largely unnecessary.  One can show that sufficient
accuracy---on the order of one percent or better---is obtained if the
average is performed only over the component
of the nucleon momentum along the beam direction. In this case one has
\begin{eqnarray}
 \left( \frac{d\sigma^{OPE+N^*}}{dM_{\pi\pi}} \right)_{av} =
\hspace{4cm}
\nonumber\\
\frac{3}{4k_F^3} \int\limits_{-k_F}^{+k_F} dk_z (k_F^2-k_z^2)
\frac{d\sigma^{OPE+N^*}}{dM_{\pi\pi}}(s_{tot}(k_z)),
\end{eqnarray}
with $s_{tot}(k_z)=m_\pi^2+M_N^2+2(\omega_{p_{lab}} E_N(k_z)-p_{lab}k_z)$
and the Fermi momentum given in terms of the local
density by $\rho_N=2k_F^3/3\pi^2$. (In fact, we adopted the same procedure
for the deuteron calculations with an `effective' Fermi momentum
representing the kinematical limit where the two nucleons recoil
together.)

Of course pions undergo substantial absorption in the nuclear medium---an 
effect which, as implied in the previous paragraph, we did not include
in the present analysis.  Here we address instead the question of 
whether the medium modifications of the final state
$\pi\pi$ interaction can account for the dramatic reshaping of the
mass distribution observed in the CHAOS experiment.
To fix the units for the in-medium results we assume that our theoretical
total cross sections account for the same fraction of the measured cross
sections as on the $^2H$ target (note again that our deuteron results
are in absolute units), which ranges between 80\% and $\sim$100\%.
The use of the in-medium $\pi\pi$ amplitudes
in the differential cross sections then  leads to the results
shown in the lower panel of figs.~\ref{+-} and \ref{++}, where we
have employed the parameterized, `effective' experimental acceptance.

In the $\pi^+\pi^+$ channel the density dependence is very weak,
the major effect being a smearing of the cross section due to the
nuclear Fermi motion. However the $\pi^+\pi^-$
\vspace{-1.5cm}
\vspace{-1.5cm}
\noindent 
channel exhibits very substantial medium effects; in particular, 
the strong threshold enhancement observed in the data is 
nicely accounted for by our model, being entirely due to the 
effects in the in-medium isoscalar $S$-wave $\pi\pi$ amplitude 
${\cal M}_{\pi\pi}^{00}$. The average nuclear density which seems 
to reproduce the $^{40}Ca$ data best turns out to be slightly above half 
the saturation density ($\rho$ = 0.5--0.7$\rho_0$).
This is in line with the average density of $^{40}Ca$ and the fact that 
the two pions are probably created near the surface. One should, however, 
realize that the two outgoing pions are at rather low energies and
can therefore pass the nucleus with little attenuation.  
Also note that the ratio of the 
threshold peak to the main maximum at $M_{\pi\pi}\simeq 350$~MeV
increases within rather small increases of density, a trend that is 
also clearly seen in the experimental data when going to heavier 
nuclei (from $^{12}C$ to $^{40}Ca$ to $^{208}Pb$)~\cite{chaos98}, 
where one would expect to probe slightly increasing average densities.

 
To summarize, we have analyzed $A(\pi^+,\pi^+\pi^\pm)$ production 
cross sections on the deuteron and nuclei at low energies using  
a realistic model of the $\pi\pi$ interaction which 
accurately describes free $\pi\pi$ scattering over a wide range 
of energies and incorporates `minimal chiral constraints'.  
Our emphasis was on the 
$S$-wave  $\pi\pi$ final state interaction and, in particular,
its modifications by the medium. The latter were included in terms of the 
standard renormalization of pion propagation through nucleon/$\Delta$-hole
excitations entering a Lippmann-Schwinger-type equation for the 
$\pi\pi$ amplitude. Many aspects of the complicated reaction 
dynamics have been treated approximately---e.g. the pion production 
process, the nuclear Fermi motion and absorption---or neglected 
altogether---e.g. Coulomb interactions and the dependence of 
the medium corrections on the total two-pion three-momentum. 
Nevertheless we find that our calculations, 
including experimental acceptance, agree reasonably well with 
recent data taken by the CHAOS collaboration at TRIUMF in both 
the $\pi^+\pi^-$ and the $\pi^+\pi^+$ channel. In particular, the 
strong enhancement of the $\pi^+\pi^-$ cross section very close to 
the free two-pion threshold can be well reproduced by  
ascribing it to the medium-modified scalar-isoscalar 
$\pi\pi$ interaction. Such an enhancement is absent in the   
isotensor $\pi^+\pi^+$ interaction, which is also consistent  
with the data. The physical origin of these features is rather simple: 
the softening of the in-medium single-pion dispersion relation acts
coherently with the attractive interaction in the isoscalar channel, 
but is not effective in the purely repulsive isotensor channel.  

The outgoing two pions being on-mass-shell precludes 
any statement as to whether the threshold enhancement of the 
in-medium isoscalar $\pi\pi$ amplitude entails substantial subthreshold 
effects at higher densities, as predicted in various $\pi\pi$ models.
However, since the $\pi^+\pi^-$ correlations in the $J$=$I$=0 channel are 
of great importance for the understanding of nuclear binding, further 
experimental and theoretical studies of the subject will certainly
be of great interest.  

\vspace{0.5cm}
 
\centerline{\bf ACKNOWLEDGMENTS}

Useful discussions with N. Grion are gratefully acknowledged. 
Two of us (JWD and RR) wish to thank Prof. J.~Speth 
for his hospitality and support during their visits to the 
Forschungszentrum J\"ulich.  One of us (RR) acknowledges 
support from the Alexander-von-Humboldt foundation as a Feodor-Lynen 
fellow, and also thanks Prof. J. Wambach for his hospitality and support 
during the visits to TU Darmstadt. 
This work was supported in part  by the U.S. Department 
of Energy under Grant No. DE-FG02-88ER40388.

\end{document}